\documentclass[%
reprint,
superscriptaddress,
nofootinbib,
twocolumn,
amsmath,amssymb,
aps,
prb,
]{revtex4-2}

\usepackage[colorlinks = true,
linkcolor = red,
urlcolor= red,
citecolor = blue,
anchorcolor = blue]{hyperref}

\usepackage{epsfig}
\usepackage{amsmath,amssymb}
\usepackage{graphicx}
\usepackage{epstopdf}
\usepackage[dvipsnames,usenames]{xcolor}
\usepackage[normalem]{ulem}
\usepackage{todonotes}
\usepackage{soul} 
\usepackage{orcidlink}
\usepackage{glossaries}
\usepackage{braket} 
\usepackage{todonotes}
\usepackage[margin=1in]{geometry}
\usepackage{ragged2e}

\usepackage{glossaries}
\newacronym{CDW}{CDW}{charge-density-wave}
\newacronym{1D}{1D}{one-dimensional}
\newacronym{2D}{2D}{two-dimensional}
\newacronym{eph}{$e$-ph}{electron-phonon}
\newacronym{DQMC}{DQMC}{determinant quantum Monte Carlo}
\newacronym{SSH}{SSH}{Su-Schrieffer-Heeger}
\newacronym{HMC}{HMC}{hybrid Monte Carlo}
\newacronym{QMC}{QMC}{quantum Monte Carlo}
\newacronym{SC}{SC}{superconducting}
\newacronym{QHO}{QHO}{quantum harmonic oscillators}
\newacronym{ML}{ML}{machine learning}
\newacronym{LBC}{LBC}{learning by confusion}
\newacronym{CNN}{CNN}{convolutional neural network}
\newacronym{PCA}{PCA}{principal component analysis}
\newacronym{t-SNE}{t-SNE}{t-distributed stochastic neighbor embedding}
\newacronym{FSS}{FSS}{finite-size scaling}

\begin{document}

\title{Charge Order in the half-filled bond-Holstein Model}

\author{Charles Jordan}
\affiliation{Department of Physics and Astronomy, University of California, Davis,
California 95616, USA}
\author{George Issa\orcidlink{0009-0009-3264-7005}}
\affiliation{Department of Physics and Astronomy, University of California, Davis,
California 95616, USA}
\author{Ehsan~Khatami\orcidlink{0000-0003-4256-6232}}
\affiliation{Department of Physics and Astronomy, San Jos\'{e} State University, San Jos\'{e}, CA 95192 USA}
\author{Richard~Scalettar\orcidlink{0000-0002-0521-3692}}
\affiliation{Department of Physics and Astronomy, University of California, Davis,
California 95616, USA}
\author{Benjamin~Cohen-Stead\orcidlink{0000-0002-7915-6280}}
\affiliation{Department of Physics and Astronomy, The University of Tennessee, Knoxville, Tennessee 37996, USA}
\affiliation{Institute for Advanced Materials and Manufacturing, The University of Tennessee, Knoxville, Tennessee 37996, USA\looseness=-1}
\author{Steven~Johnston\orcidlink{0000-0002-2343-0113}}
\affiliation{Department of Physics and Astronomy, The University of Tennessee, Knoxville, Tennessee 37996, USA}
\affiliation{Institute for Advanced Materials and Manufacturing, The University of Tennessee, Knoxville, Tennessee 37996, USA\looseness=-1}



\begin{abstract}
We use determinant quantum Monte Carlo to study the half-filled `bond-Holstein' model on a square lattice. We find that the model exhibits a \gls*{CDW} phase transition with a critical temperature $T_\mathrm{cdw}$ considerably higher than that of the canonical `site-Holstein' model. Using a finite-size scaling analysis of the charge structure factor $S_{\rm cdw}$, we obtain $T_\mathrm{cdw}$ to greater than one percent accuracy. At the same time, local observables also show clear signatures consistent with the transition temperatures inferred from our scaling analysis. We attribute the enhanced \gls*{CDW} tendencies to a phonon-mediated nearest-neighbor electron repulsion that is directly proportional to the dimensionless electron-phonon coupling $\lambda$ in the atomic ($t\rightarrow 0$) limit. This behavior contrasts with the site-Holstein case, where the same limit yields only an on-site attraction. We supplement our analysis with results from several unsupervised machine learning methods, which not only confirm our estimates of $T_\mathrm{cdw}$ but also provide insight into the high-temperature crossover between a metallic and bipolaron liquid regime.
\end{abstract}

\maketitle

\glsresetall

\section{Introduction}
The Holstein Hamiltonian~\cite{Holstein1959studiesI, Holstein1959studiesII}, in which atomic  displacements couple to their local charge density, is a widely studied model of \gls*{eph} interactions. It has been used to describe both small polaron formation in the dilute limit (density $\langle n \rangle \ll 1$)~\cite{DeRaedt1983numerical, Bonica1999Holstein, Berciu2006green}, and competition between superconductivity and \gls*{CDW} order at higher densities~\cite{hirsch82, scalettar89, marsiglio90, Vekic1992, Berger1995, Huang2003, Hohenadler2004, dee2019temperature, esterlis18, cohenstead20, nosarzewski21, Bradley2023}. At half-filling and on a bipartite lattice, the \gls*{CDW} states involve the formation of empty and doubly occupied sites alternating on the two sublattices. Upon doping away from half-filling ($0 < \langle n \rangle < 1$), either conventional ($s$-wave) \gls*{SC} states~\cite{dee2019temperature, bradley21, nosarzewski21} or (bi)polaron liquid-like~\cite{esterlis18, nosarzewski21, issa2025learning} states can form, depending on the strength of the \gls*{eph} coupling, phonon energy, and filling. These phases are driven by an effective phonon-mediated onsite attraction between spin-up and spin-down electrons, which leads to a local bipolaron formation (pairing of fermions). At low-to-intermediate densities, these pairs condense into a \gls*{SC} phase but closer to commensurate fillings, numerical studies indicate that bipolarons tend to phase separate rather than superconduct~\cite{nosarzewski21}.

\begin{figure}[b]
    \centering
    \includegraphics[width=\columnwidth]{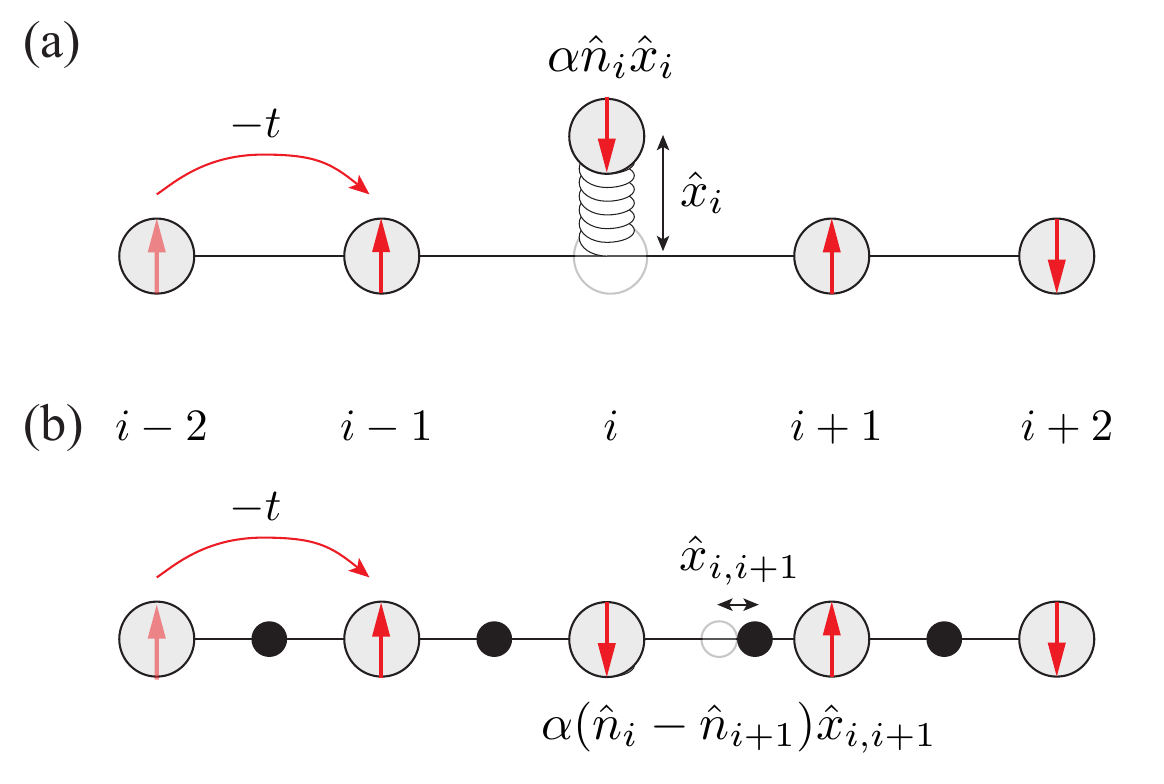}
    \caption{A cartoon sketch of the (a) site- and (b) bond-Holstein models (see also Sec.~\ref {subsec:Holstein}), shown here in one dimension for simplicity. In the site-Holstein model, local lattice displacements couple to the total local electron charge density, leading to a momentum independent \gls*{eph} interaction. In the bond-Holstein model, generalized oscillators are defined on each of the system's bonds and couple to neighboring electron densities with opposite signs. This mechanism results in a momentum dependent \gls*{eph} coupling $g(\boldsymbol{q})$.}\label{fig:models}
\end{figure}

\Gls*{QMC} methods have played a key role in studying the canonical Holstein model [which we refer to here as the ``site-Holstein model,'' see Fig.~\ref{fig:models}(a)]
~\cite{peierls79, hirsch82, hirsch83, scalettar89, marsiglio90,freericks93}. But it has only been relatively recently (the last 5-10 years) that the critical temperature $T_\mathrm{cdw}$ has been determined with high precision on various lattice geometries
~\cite{ohgoe17, weber2018, zhang2019, hohenadler19, feng2020, bradley21, nosarzewski21, araujo22}. For the square lattice, \gls*{CDW} order forms at half filling at low temperature for all values of the \gls*{eph} coupling $\lambda$ owing to the divergence of the non-interacting density of states~\cite{hirsch82, scalettar89, marsiglio90}. For the honeycomb lattice, where the density of states at half-filling vanishes, a critical coupling $\lambda_\mathrm{c}$ separates a $T=0$ semi-metal from the \gls*{CDW} phase~\cite{Bradley2023}. This geometry-dependent phenomenology is analogous to the behavior of antiferromagnetic order in the repulsive Hubbard model for these two lattices~\cite{white89, sorella1992semi, paiva2005ground}.

The site-Holstein model is the simplest description of \gls*{eph} coupled systems. More recently, there has been growing interest in studying more complicated (and realistic) models. For example, \gls*{QMC} has been recently applied to studying variants of the \gls*{SSH} model~\cite{Barisic1970tightbinding, su1980soliton} in which the phonons couple to the fermionic hopping between sites~\cite{li2020quantum, xing2021quantum, malkaruge2023comparative, TanjaroonLy2023, cai2021antiferromagnetism}. This article explores a case that is intermediate between the site-Holstein and \gls*{SSH} models, which we refer to as the ``bond-Holstein model.'' It is similar to the Holstein model in that the phonons couple to the charge density rather than the hopping, but resembles the \gls*{SSH} model in that the phonon degrees of freedom live on the lattice bonds and couple to the neighboring electron densities [see Fig.~\ref{fig:models}(b)]. 

Here, we study the half-filled bond-Holstein model using a combination of numerically exact \gls*{DQMC} simulations and several machine-learning methods to map its ground-state phase diagram. We find that the model hosts enhanced \gls*{CDW} correlations and bipolaronic tendencies compared to the canonical site-Holstein model, with a transition temperature $T_{\rm cdw}$ that remains large and nonzero in the atomic limit. This model has been proposed as a low-energy single-band model for transition metal oxides like the cuprates or bismuthates, where the bond-stretching motion of negatively charged oxygen anions directly modulates the site energies of the neighboring cation sites~\cite{Song1995electron, Song1995erratum}. Our results are thus relevant for understanding various charge orders appearing in these materials~\cite{Sleight2015bismuthates, Arpaia2021charge}.

This paper is organized as follows: Section~\ref{sec:model+methods} defines the bond-Holstein Hamiltonian and discusses its similarities and differences with the site-Holstein model. This section also provides some preliminary discussion of the bond model's atomic ($t=0$) limit and the details of our \gls*{DQMC} simulations. With this background, Sec.~\ref{sec:results} presents our simulation results for the \gls*{CDW} transition and several local observables at half-filling that encode information about the transition. Sec.~\ref{sec:results_DQMC} shows results for our \gls*{DQMC} simulations of the full model while Sec.~\ref{sec:results_atomic} analyzes the model's (nontrivial) atomic limit. Next, Sec.~\ref{sec:lbc} applies a suite of unsupervised \gls*{ML} approaches, including \gls*{PCA}, \gls*{t-SNE}, and \gls*{LBC}, to \gls*{DQMC} configurational snapshots to provide further insight into the physics. In particular, this section focuses on identifying high temperature regimes where bipolaron formation occurs without establishing long-range \gls*{CDW} order. These results culminate in a phase diagram presented in Sec.~\ref{sec:phase_diagram}. Finally, Sec.~\ref{sec:conclusions} summarizes our results and discusses their relevance to materials.

\section{Model and methods}\label{sec:model+methods}
\subsection{Holstein Hamiltonian Variants}\label{subsec:Holstein}

The Hamiltonians for the two variants of the Holstein model we consider here are partitioned as
\begin{equation}\label{eq:H_site}
\hat{H}_{\rm site} = \hat{K} + \hat{U}_{\rm site} + \hat{V}_{\rm site}
\end{equation}
and
\begin{equation}\label{eq:H_bond}
\hat{H}_{\rm bond} = \hat{K} + \hat{U}_{\rm bond} + \hat{V}_{\rm bond}.
\end{equation}

The models share the same electron kinetic energy term $\hat{K}$, which describes electron hopping between nearest-neighbor sites
\begin{equation}\label{eq:hopping}
\hat{K} = -t \sum_{\langle i,j\rangle,\sigma}
\left(\hat{c}^\dagger_{i,\sigma}\hat{c}^{\phantom{\dagger}}_{j\sigma} + \mathrm{h.c.}\right)
-\mu \sum_{i,\sigma} \hat n_{i,\sigma}.
\end{equation}
Here $\hat{c}^\dagger_{i,\sigma}$ and $\hat{c}^{\phantom{\dagger}}_{i,\sigma}$ are fermion creation and annihilation operators for site $i$ and spin $\sigma$, $\hat n_{i,\sigma} = \hat{c}^\dagger_{i,\sigma} \, \hat{c}^{\phantom{\dagger}}_{i,\sigma}$ is the associated number operator, $\langle i,j\rangle$ denotes a sum over nearest neighbor sites, $t$ is the nearest-neighbor hopping integral, and $\mu$ is the chemical potential, which controls the band filling.

Equations \eqref{eq:H_site} and \eqref{eq:H_bond} differ in how they treat the lattice ($\hat{U}$) and \gls*{eph} ($\hat{V}$) interactions. The site model, which corresponds to the traditional Holstein Hamiltonian~\cite{Holstein1959studiesI, Holstein1959studiesII}, introduces local \gls*{QHO} on every site that couple to the local electron density
\begin{equation}\label{eq:lattice_site}
\begin{split}
\hat{U}_{\rm site} &= \sum_i
\frac{m \omega_0^2}{2} \, \hat{x}_i^2
+\sum_i \frac{1}{2m} \, \hat{p}_i^2,\\
\hat{V}_{\rm site} &= \alpha \sum_{i,\sigma} \hat{x}_i \, \left( \hat{n}_{i,\sigma} -\frac{1}{2} \right).
\end{split}
\end{equation}
Here, $\hat{x}_i$ and $\hat{p}_i$ are the position and momentum operators for the atom at site $i$, $m$ is its mass, $\omega_0$ is the frequency of the oscillator, and $\alpha$ is the strength of the \gls*{eph} coupling. 

The bond model introduces \glspl*{QHO} on each nearest-neighbor bond with generalized position $\hat{x}_{\langle i,j\rangle}$ and momentum $\hat{p}_{\langle i,j\rangle}$ operators, and couples the bond displacement to the neighboring charge densities
\begin{equation}\label{eq:lattice_bond}
\begin{split}
\hat{U}_{\rm bond} &= \sum_{\langle i,j\rangle}
\frac{m \omega_0^2}{2} \, \hat{x}_{\langle i,j\rangle}^2
+ \sum_{\langle i,j\rangle}\frac{1}{2m} \, \hat{p}_{\langle i,j\rangle}^2, \\
\hat{V}_{\rm bond} &= \alpha \sum_{{\langle i,j\rangle},\sigma} \hat{x}_{\langle i,j\rangle} \, \big( \hat{n}_{i,\sigma} -\hat{n}_{j\sigma} \big).
\end{split}
\end{equation}
In this case, $m$ and $\omega_0$ are understood to be the effective mass and
frequency of the bond oscillator, respectively. In this case, $\hat{x}_{\langle i,j\rangle}$ and $\hat{p}_{\langle i,j\rangle}$ describe the dynamics of an individual bond between sites $i$ and $j$; they should not be read as a difference between atomic coordinates, i.e., $\hat{x}_{\langle i,j\rangle}\ne\hat{x}_i - \hat{x}_j$.

The phonon modes in both models are similar in that they describe a collection of independent (hence dispersionless) \glspl*{QHO}. The distinction between the models is that the bond-Holstein model introduces longer-range coupling between the two sites terminating each bond. This model is sometimes  regarded as a single-band effective model for bond-stretching modes in materials such as transition-metal oxides~\cite{Song1995electron, Slezak2006, Lau2007single}. For example, one can interpret the bond phonons as describing the atomic motion of the negatively charged anions in the system, which modulates the on-site orbital energies of neighboring cations via an electrostatic coupling.

Throughout this work, we set $m=\hbar =t=1$ and $\omega_0 = t$, and solve Eqs.~\eqref{eq:H_site} and \eqref{eq:H_bond} on a half-filled square lattice of linear dimension $L$ (volume $N=L^2$) with periodic boundary conditions. The \gls*{eph} interaction terms for both models are particle-hole symmetric so that half-filling $\langle n \rangle=\frac{1}{N}\sum_{i,\sigma}\langle \, \hat n_{i,\sigma} \, \rangle = 1$ occurs at $\mu=0$, where also $\langle \hat{x}_i \rangle$ and $\langle \hat{x}_{i,j} \rangle = 0$. The bare electron dispersion relation is given by $\epsilon(\boldsymbol{k})=-2\,t \, [\cos(k_xa)+\cos(k_ya)]-\mu$, where $a$ is the lattice parameter, with a corresponding bandwidth $W=8t$. Finally, we adopt the dimensionless \gls*{eph} coupling constants $\lambda_\text{site}=\alpha^2/(\omega_0^2W)$ for the site model and $\lambda_\text{bond} = 4\alpha^2/(\omega_0^2W)$ for the bond model~\cite{Slezak2006, Lau2007single}. The factor of $4$ in the bond definition reflects the fact that an electron on a given site interacts with $z=4$ neighboring phonon modes.

\subsection{The atomic limit ($t=0$)}\label{sec:model_atomic}
One can obtain a useful intuition into the physics of the site-Holstein model by setting the hopping $t=0$. In this limit, one can then complete the square in the phonon position to obtain an on-site phonon-mediated attractive interaction $U_{\text{eff}} =-\alpha^2/\omega_0^2$ between electrons of opposite spin. This attraction causes local, on-site pair (bipolaron) formation. These pairs can then either order spatially or condense into an $s$-wave superconducting phase when the hopping is restored. The former phase is most dominant at half-filling on a bipartite lattice~\cite{marsiglio90, scalettar89, bradley21, nosarzewski21}, since pairs and empty sites can each occupy their own sublattice. Maximizing the number of empty sites around a doubly occupied site optimizes the virtual hopping away from the pair, and hence lowers the energy. This drives the formation of a $\boldsymbol{Q} = (\pi,\pi)/a$ \gls*{CDW} order on a \gls*{2D} square lattice at half-filling. It is important to note that this energy lowering is reliant on a second-order perturbation in $t$. For the site-Holstein model, $T_{\rm cdw}/t \sim 0.2$ for commonly studied values of $\alpha$ and $\omega_0$~\cite{marsiglio90, scalettar89, esterlis18, dee2019temperature, bradley21, nosarzewski21}. On the other hand, the superconducting transition occurs at much lower temperatures ($T_{\rm sc}/t \sim 0.03-0.05$)~\cite{bradley21, nosarzewski21} and is maximized at intermediate densities $\langle n \rangle \sim 0.5-0.85$.\footnote{The precise values depend on the phonon energy $\omega_0$, and significant superconducting correlations have largely only been successfully resolved in models $\omega_0/t \gtrapprox 1$.}

We can carry out a similar analysis for the {\it bond}-Holstein model. The resulting effective electron-electron interaction is
\begin{equation}\label{eq:Veff}
\hat{V}_\mathrm{eff} = -\frac{|U_{\rm eff}|}{2} \sum_{\langle i,j\rangle}\left(\hat{n}_{i} - \hat{n}_{j} \right)^2,
\end{equation}
where
$\hat{n}_{j} = \sum_\sigma \hat{n}_{j,\sigma}$
is the total electron density on site $i$. This effective phonon-mediated electron-electron interaction can also be derived by performing a Lang-Firsov transformation~\cite{Slezak2006, Lau2007single}. Eq.~\eqref{eq:Veff} has both an on-site attraction between spin up and spin down electrons and an {\it inter}-site repulsion between fermions of all spin labels. As a consequence, \gls*{CDW} physics is not reliant upon the hopping $t$ but naturally arises in the $t=0$ limit. We can thus expect that the bond model's transition temperature and its scaling with Hamiltonian parameters will differ significantly from those of the site model. Our \gls*{DQMC} simulations will confirm and quantify this difference. We will also show classical Monte Carlo results for the $t=0$ limit.

\subsection{Momentum dependence of the coupling}\label{sec:model_gq}
Another useful way to distinguish the site- and bond-Holstein models is by re-writing the \gls*{eph} interaction in momentum space.

In the site-Holstein case, the interaction can be expressed as
\begin{eqnarray}
\hat V_{\rm site} = \frac{g}{\sqrt{N}} \sum_{\boldsymbol{k},\boldsymbol{q},\sigma}\hat c^{\dagger}_{\boldsymbol{k}+\boldsymbol{q},\sigma} \hat c^{\phantom{\dagger}}_{\boldsymbol{k},\sigma}\big( \hat b^{\dagger}_{-\boldsymbol{q}}+ \hat b^{\phantom{\dagger}}_{\boldsymbol{q}}\big),
\label{eq:siteKspace}
\end{eqnarray}
where $g= \alpha / \sqrt{2 m \omega_0}$. Notice here that the \gls*{eph} coupling 
is independent of both the electron ($\boldsymbol{k}$) and phonon ($\boldsymbol{q}$) momenta, which is a consequence of all operators acting on the same site. In contrast, the interaction in the momentum space for the bond model takes the form~\cite{Slezak2006}
\begin{eqnarray}
\hat V_{\rm bond} = \frac{1}{\sqrt{N}} \sum_{\boldsymbol{k},\boldsymbol{q},\sigma} g(\boldsymbol{q})
\hat c^{\dagger}_{\boldsymbol{k}+\boldsymbol{q},\sigma} \hat c^{\phantom{\dagger}}_{\boldsymbol{k},\sigma}\big(\hat b^{\dagger}_{-\boldsymbol{q}} +\hat b^{\phantom{\dagger}}_{\boldsymbol{q}}\big).
\label{eq:bondKspace}
\end{eqnarray}
Here, $g(\boldsymbol{q}) = \big( \alpha / \sqrt{2 m \omega_0} \, \big)\sum_\nu 2\mathrm{i}\sin\left(q_\nu a/2\right)$ and the sum over $\nu$ runs over the spatial dimensions. The key point is that placing the phonons on the bonds of the lattice and coupling them to the neighboring electron density results in an \gls*{eph} vertex that now depends explicitly on the phonon momentum $\boldsymbol{q}$. Indeed, $g(\boldsymbol{q})$ is maximal at $\boldsymbol{q}=(\pi,\pi)/a$ in 2D, which further indicates the strong \gls*{CDW} response at that momentum, and vanishes at $\boldsymbol{q}=0$. A similar momentum dependence has been derived for the \gls*{eph} coupling to Cu-O and Bi-O bond-stretching phonons in multiorbital models for the cuprates~\cite{Rosch2004electron, Johnston2010systematic} and bismuthates~\cite{CohenStead2023hybrid}. This similarity further underscores the relevance of the bond-Holstein model for those materials.

\subsection{Determinant Quantum Monte Carlo}\label{sec:DQMC}

We solve Eqs.~\eqref{eq:H_site} and \eqref{eq:H_bond} using numerically exact \gls*{DQMC} with combined \gls*{HMC} and global swap updates for the phonon fields, as implemented in the \texttt{SmoQyDQMC.jl} package~\cite{SmoqyDQMC1, SmoqyDQMC2}. Throughout, we consider \gls*{2D} square lattices of linear dimension $L$ and adopt an imaginary time discretization of $\Delta\tau = 0.05/t$. We perform 100 fermionic time steps in the \gls*{HMC} update, 5,000 burn in updates to thermalize the system, and then 10,000 measurement updates once the system is thermalized. To accelerate the thermalization process, we initialize the phonon configurations with the expected $\boldsymbol{q}=(\pi,\pi)/a$ displacement pattern.

\begin{figure}[t]
\centerline{
\includegraphics[width = 1.1\columnwidth]{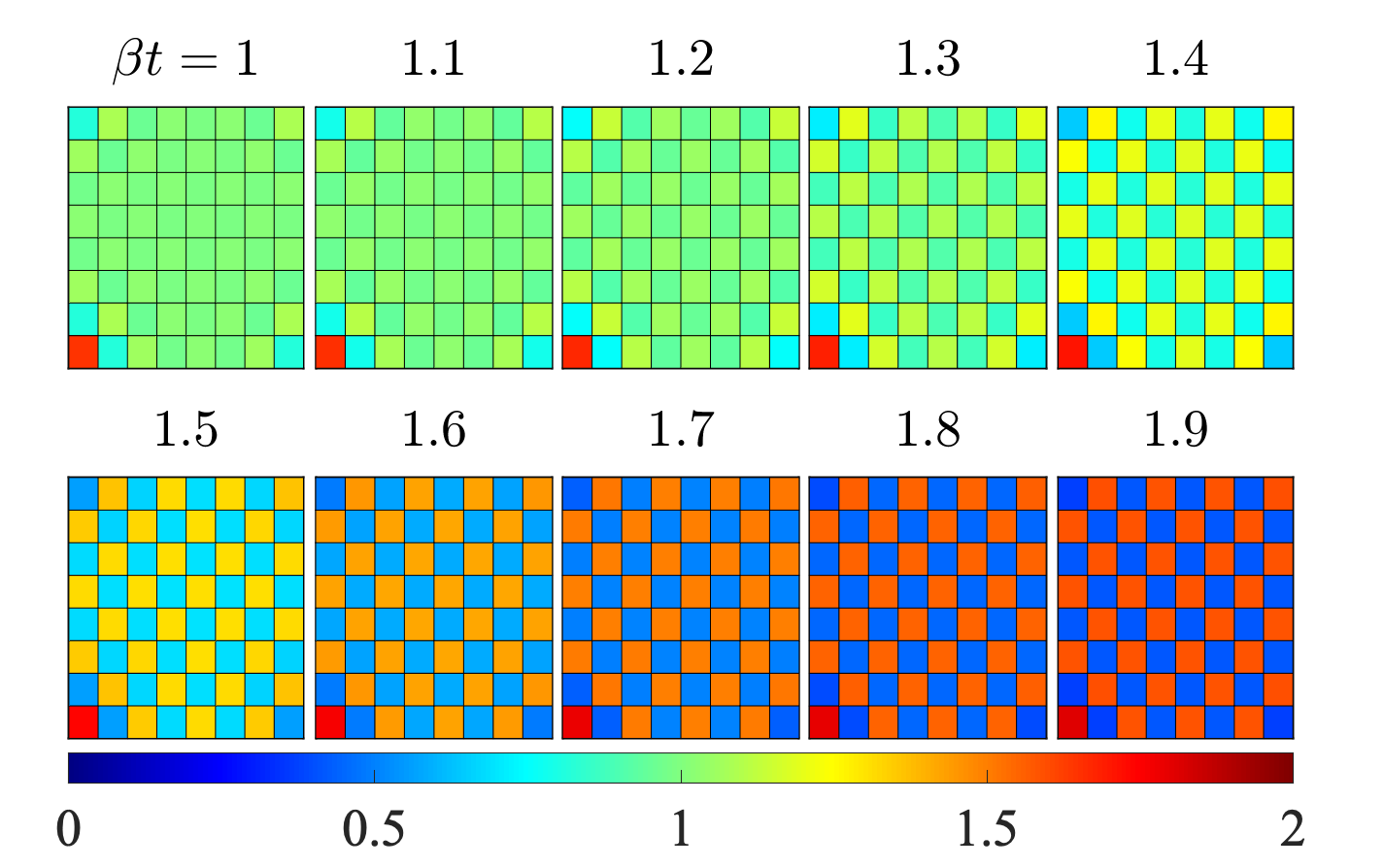}}
\caption{ Position space snapshots of the charge density correlation function $\langle \hat{n}_i \hat{n}_0 \rangle$ in the bond-Holstein model as a function of $\beta$, obtained on an $L=8$ lattice with dimensionless coupling $\lambda_{\rm bond}=0.4$. The strength of the correlations at each lattice site relative to the reference site in the bottom left is indicated by the common color scale at the bottom, where red (blue) corresponds to high (low) correlation.}\label{fig:CDWsnaps}
\end{figure}

To assess the strength of the \gls*{CDW} correlations, we measure the charge density structure factor
\begin{equation}\label{eq:Scdw}
S_{\rm cdw}(\boldsymbol{q}) = \frac{1}{N} \sum_{i,j} \langle \hat{n}_i \hat{n}_j \rangle \, e^{-\mathrm{i}\boldsymbol{q}\cdot(\boldsymbol{R}_i-\boldsymbol{R}_j)},
\end{equation}
where $\boldsymbol{R}_i$ is the lattice vector for site $i$. At half-filling, the dominant correlations appear at $\boldsymbol{q}=(\pi,\pi)/a$, and so we focus on this momentum point throughout the text. At high temperatures, the real-space correlations $ \langle \hat{n}_i \hat{n}_j \rangle $ are short ranged and the double sum over all pairs of nearest-neighbor sites $i,j$ is proportional to $N$. Thus, for our choice of normalization, $S_{\rm cdw}(\pi,\pi)$ will be roughly independent of lattice size in this regime. On the other hand, in the low-temperature ordered phase, $ \langle \hat{n}_i \hat{n}_j \rangle $ develops long-range modulations at particular $\boldsymbol{q}$ vectors. For example, Fig.~\ref{fig:CDWsnaps} shows the evolution of $\langle \hat{n}_i \hat{n}_j \rangle$ for the half-filled bond model as a function of temperature, where one sees the formation of a $\boldsymbol{q}=(\pi,\pi)/a$ modulation as the temperature decreases. In this regime, the double sum over all pairs of sites $i,j$ in Eq.~\eqref{eq:Scdw} is proportional to $N^2$ and $S_{\rm cdw}(\boldsymbol{q})$ becomes an extensive quantity, growing in proportion to $N$.

We also measure several local observables as a function of model parameters. These include the average electron kinetic energy per site,
\begin{equation}\label{eq:E_kin}
\mathcal{K} = -\frac{t}{N} \sum_{\langle i,j\rangle,\sigma}
\left\langle \hat{c}^\dagger_{i,\sigma}\hat{c}^{\phantom{\dagger}}_{j,\sigma} + \text{h.c.} \right\rangle,
\end{equation}
the average \gls*{eph} energy per site
\begin{equation}
{\cal V} = \frac{\alpha}{N} \sum_{\langle i,j \rangle,\sigma}
\left\langle \hat{x}_{\langle ij\rangle} \left( \hat{n}_{i,\sigma} -\hat{n}_{j,\sigma}\right) \right\rangle,
\end{equation}
and the double occupancy per site
\begin{equation}\label{eq:D}
{\cal D} = \frac{1}{N}\sum_i\langle \hat{n}_{i,\uparrow} \hat{n}_{i,\downarrow} \rangle.
\end{equation}
These quantities, being derived from local observables, are often less susceptible to finite-size effects and generally converge to the thermodynamic limit for modest lattice sizes.

\section{Monte Carlo Results}\label{sec:results}
\subsection{DQMC Simulations}\label{sec:results_DQMC}

We begin by examining the real-space density correlations of the bond model as a function of temperature, as shown in Fig.~\ref{fig:CDWsnaps}. Here, we show the equal-time density-density correlation function $\langle \hat{n}_i \hat{n}_0 \rangle$ as a function of distance from the origin obtained on an $L = 8$ lattice with $\lambda_\mathrm{bond} = 0.4$. At high temperatures ($\beta t \approx 1)$, the real-space correlations are flat, indicative of a uniform charge density. Lowering the temperature (increasing $\beta$) produces a robust checkerboard-like modulation, which extends across the entire cluster already for $\beta t \approx 1.4$. This behavior is evidence for the formation of a long range $\boldsymbol{q} = (\pi,\pi)/a$ \gls*{CDW} order at relatively high temperatures.

\begin{figure}[t]
\includegraphics[width=\columnwidth]{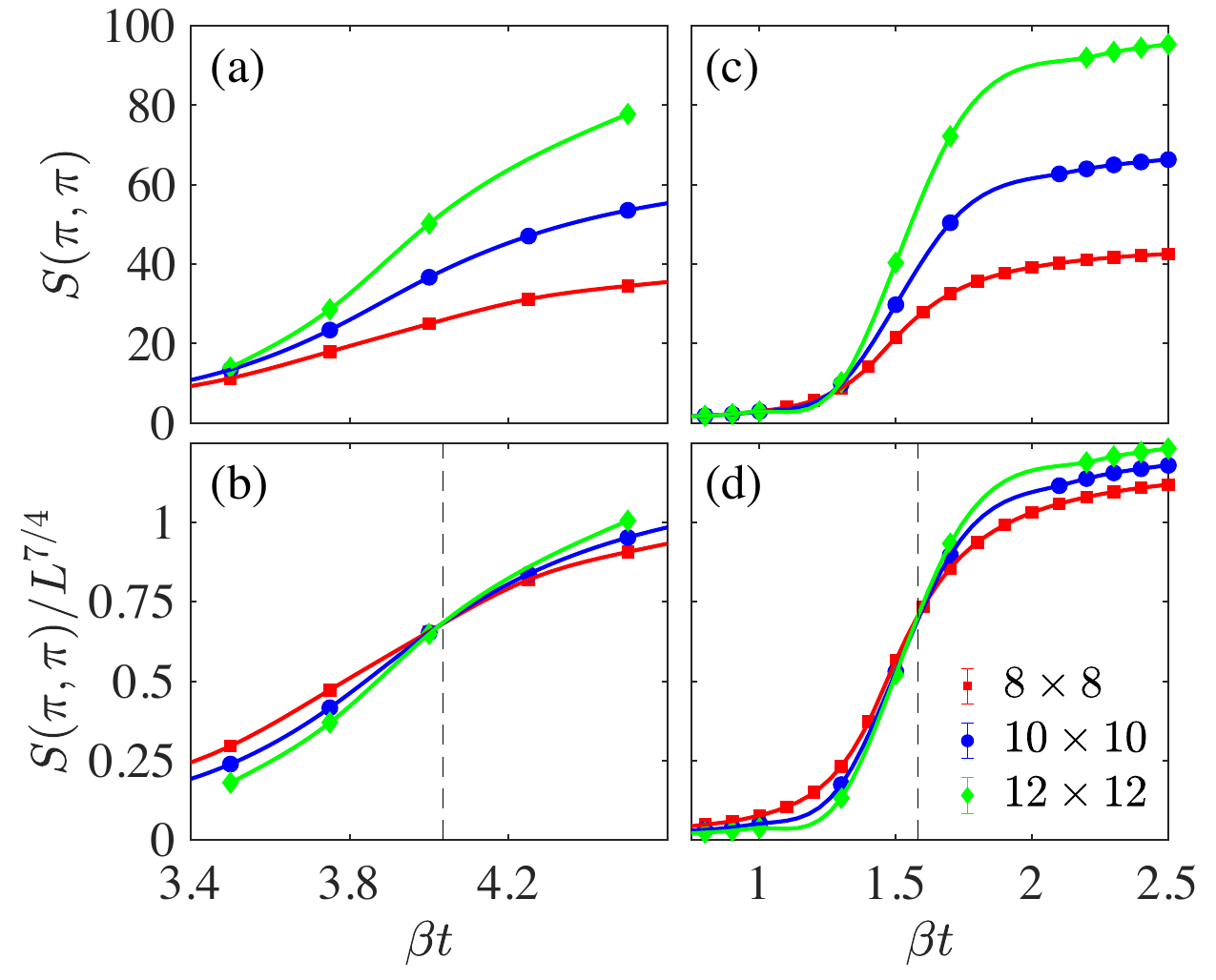}
\caption{(a) \& (b) Structure factor $S(\pi,\pi)$ and $S(\pi,\pi)/L^{\gamma/\nu}$ for the site-Holstein model with the Ising universality class exponents ${\gamma}/{\nu} = {7}/{4}$ and different lattice sizes $L$. Panels (c) \& (d) show the analogous data for the bond-Holstein model. The vertical dashed line in the lower panels indicates the transition temperature $T_{\rm cdw}$ determined from the crossing of the curves for different lattice sizes (see text). In both cases, the dimensionless coupling has been fixed to $\lambda_{\rm bond}=0.4$. The apparent difference in abruptness of the two \gls*{CDW} transitions is a consequence of the more `zoomed-in' range of $\beta$ in the site case.}
\label{fig:ScaledScdw}
\end{figure}

We can further quantify the formation of the \gls*{CDW} order by tracking the structure factor $S(\pi,\pi)$ as a function of inverse temperature $\beta$, as shown in Fig.~\ref{fig:ScaledScdw}. Here, we compare results for the site- and bond-Holstein models in the left and right columns, respectively, for three lattice sizes ($L=8$, $10$, and $12$). The top row plots the raw structure factors, which evolve from a lattice-size-independent value at small $\beta$ (high $T$) to increasing with lattice size at large $\beta$ (low $T$) for both models. As remarked earlier, this behavior is the expected trend in entering an ordered phase. The critical temperature $T_{\rm cdw}$ can be roughly inferred from the values at which this change in lattice size dependence occurs.

A more precise estimate of $T_{\rm cdw}$ can be obtained through a finite-size scaling analysis~\cite{cardy2012finite}. The \gls*{CDW} transition breaks a $\mathbb{Z}_2$ symmetry and thus falls in the \gls*{2D} Ising universality class. Making use of the fact that the structure factor diverges as $S_{\rm cdw} \sim (T-T_{\rm cdw})^{-\gamma}$ while the correlation length $\xi$ diverges as $\xi \sim (T-T_{\rm cdw})^{-\nu}$ at the critical point $T_{\rm cdw}$ in the thermodynamic limit, one can show that
\begin{equation}
S(L,T) = L^{\frac{\gamma}{\nu}} \, f\left( L [T-T_{\rm cdw}]^{\nu} \right)
\label{eq:fss}
\end{equation}
on finite-size lattices. Here, $f(x)$ is an unknown, but universal, function of its argument. Eq.~\eqref{eq:fss} is based on the fact that the physics is determined by the ratio of length scales $L/\xi \sim L (T-T_{\rm cdw})^{\nu} $. The key observation is that $ L^{-\frac{\gamma}{\nu}} S(L,T_{\rm cdw}) = f(0)$ is independent of $L$. Therefore, structure factor curves for different lattice sizes should cross at the critical temperature when scaled by $ L^{-\frac{\gamma}{\nu}} $. The exponents for the \gls*{2D} Ising universality class are $ \gamma = \frac{7}{4} $ and $\nu = 1$. A similar analysis also holds for the site-Holstein model.

\begin{figure}[t]
    \includegraphics[width = \columnwidth]{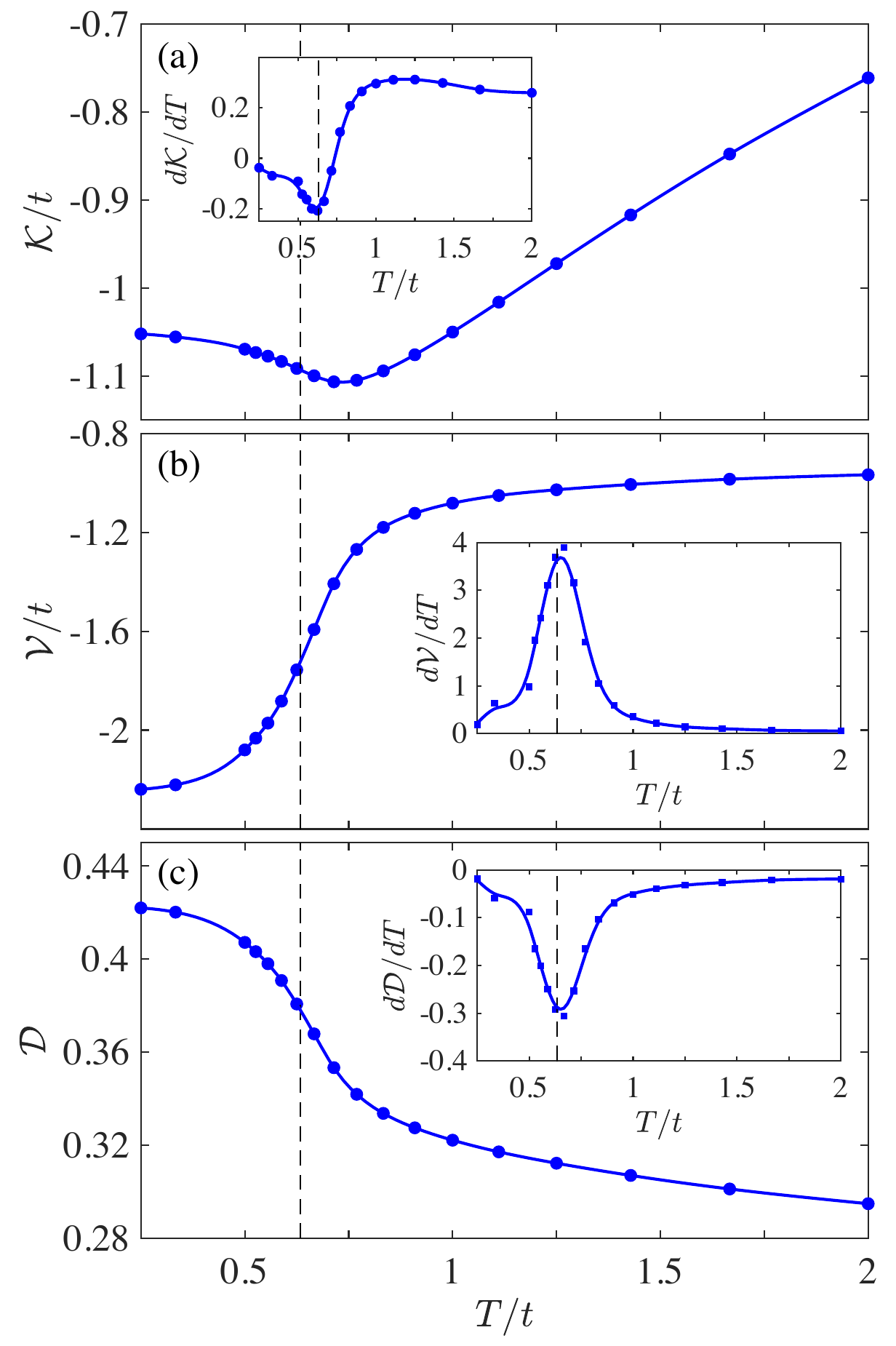}
    \caption{Average (a) electron kinetic energy $\mathcal{K}$, (b) $e$-ph potential energy $\mathcal{V}$, and (c) double occupancy $\mathcal{D}$ as functions of temperature for the bond-Holstein model with dimensionless coupling $\lambda_{\rm bond}=0.4$ and lattice size $L=8$. The insets plot the first derivative of each quantity with respect to temperature. The vertical dashed lines indicate the \gls*{CDW} transition temperature, as determined in Fig.~\ref{fig:ScaledScdw}.} \label{fig:local_observables}
\end{figure}

Applying this analysis to the structure factor data for the site [Fig.~\ref{fig:ScaledScdw}(b)] and bond-Holstein [Fig.~\ref{fig:ScaledScdw}(d)] models yield crossings at $\beta_\mathrm{cdw} \approx 4.00/t$ ($T_\mathrm{cdw} = 0.248t$) and $\beta_\mathrm{bond}\approx 1.58/t$ ($T_\mathrm{cdw} \approx 0.63$), respectively, as indicated by the black dashed lines. One key conclusion of Fig.~\ref{fig:ScaledScdw} is that $T_{\rm cdw}$ is substantially higher (about a factor of three for the parameters shown) in the bond-Holstein model relative to the site-Holstein case. As discussed earlier, we attribute this enhancement to the additional effective intersite repulsion mediated by bond phonons, which grows with \gls*{eph} coupling $\lambda$.

We now turn to the behavior of the observables given in Eqs.~\eqref{eq:E_kin}-\eqref{eq:D}. Fig.~\ref{fig:local_observables}(a)-(c) shows the temperature evolution of the electron kinetic energy ${\cal K}$, average \gls*{eph} interaction energy $\mathcal{V}$, and double occupancy $\mathcal{D}$, obtained on an $L = 8$ cluster with $\lambda_\mathrm{bond} = 0.4$. The corresponding insets show the first derivative of each quantity with respect to $T$. As the temperature $T$ is lowered, ${\cal K}$ increases in absolute value, reflecting the preferred occupation of the lower energy levels of the electron band structure. However, this trend is interrupted at $T_{\rm cdw}$ (indicated by the dashed line), where ${\cal K}$ begins to decrease in magnitude with decreasing $T$. We interpret this as a signal of entry into the insulating \gls*{CDW} phase, where the electron mobility is inhibited. A similar change in behavior can also be seen in the average \gls*{eph} energy ${\cal V}$, shown in Fig.~\ref{fig:local_observables}(b). At the \gls*{CDW} transition, ${\cal V}$ shows an abrupt increase in magnitude. ($\mathcal{V}$ is negative, reflecting the attractive nature of the \gls*{eph} interaction.) When ordering occurs, the energy is lowered further as pairs become more robust. Thus ${\cal V}$ becomes larger in magnitude. The peak in $\frac{d{\cal V}}{dT}$ aligns almost perfectly with the $S_{\rm cdw}$ crossing (vertical dashed lines).

Finally, the double occupancy, ${\cal D}$ in Fig.~\ref{fig:local_observables}(c) grows rapidly at $T_{\rm cdw}$ and mirrors the evolution of ${\cal V}$. At high temperature, the average value of the double occupancy approaches the value expected for a noninteracting system, ${\cal D}=\langle n_{\uparrow} n_{\downarrow} \rangle =\langle n_{\uparrow} \rangle \langle n_{\downarrow} \rangle = (1/2)^2$. Conversely, there is a significant enhancement of the double occupancy at low temperatures where \gls*{CDW} ordering occurs, consistent with the electrons forming pairs on one of the two sublattices.

\begin{figure}[t!]
\includegraphics[width =\columnwidth, angle=0]{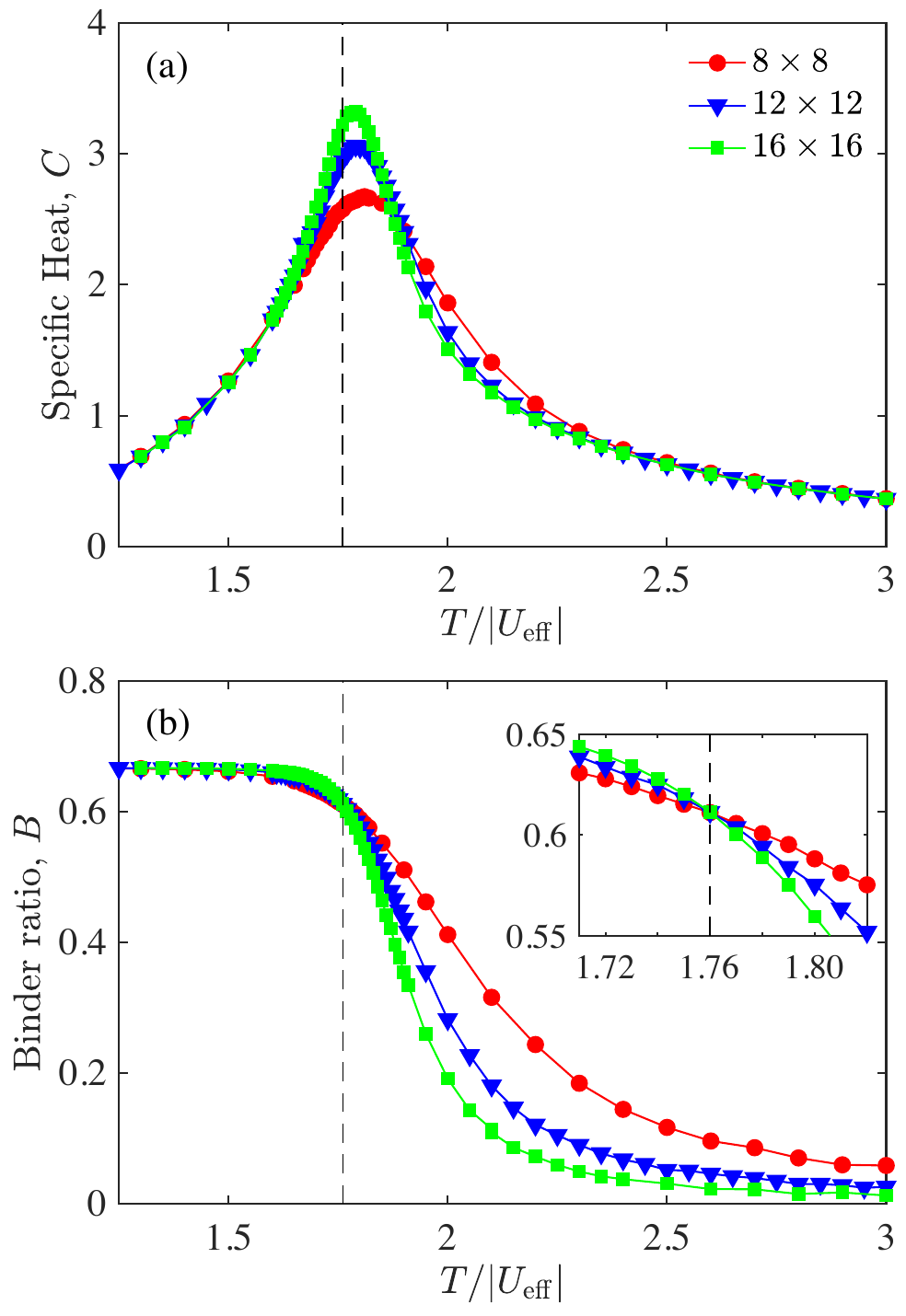}
\caption{(a) Specific heat $C$ of the bond-Holstein model as a function of temperature.
(b) The Binder ratio $B$ (see text) as a function of temperature. The vertical dashed lines in both panels denote the position of the Binder ratio crossing derived from panel (b). The legend in panel (a) is common to both panels. $|U_\mathrm{eff}| = 1$ in these simulations.
}
\label{fig:teq0}
\end{figure}

\subsection{The atomic ($t=0$) limit}\label{sec:results_atomic}
As noted in Sec.~\ref{sec:model_atomic}, the bond-Holstein model has a nontrivial $t=0$ limit in which electrons on adjacent sites experience a repulsion set by the scale $|U_\mathrm{eff}|=\alpha^2/\omega_0^2$. This behavior contrasts with the site-Holstein model, where the sites decouple in the absence of fermionic hopping. Since quantum charge fluctuations from the hopping processes have been turned off in this limit, we expect that the bond-Holstein model will have nonzero \gls*{CDW} critical temperature while $T_\text{cdw} = 0$ for the site-Holstein model.

Figure~\ref{fig:teq0} shows the results of classical Monte Carlo simulations of the bond model in the atomic limit. ($T$ in this case is measured in units of $|U_{\rm eff}|$, the sole energy scale in this situation.) 
Fig.~\ref{fig:teq0}(a) plots the specific heat $C$ obtained from simulations of different lattice sizes $L = 8$, $12$, and $16$. For all lattice sizes, it exhibits a peak that sharpens and increases in height as the system size increases. The location of this peak in the thermodynamic limit is indicative of the \gls*{CDW} phase transition. A precise determination of $T_{\rm cdw}$ can be obtained by the crossing of the Binder ratio $B = 1 - \langle M_{\rm cdw}^4 \rangle / (3 \, \langle M_{\rm cdw}^2 \rangle^2)$, 
where $M_{\rm cdw}$ is the difference of the sums of the fermionic occupations 
on the two sublattices of the bipartite square geometry. Fig.~\ref{fig:teq0}(b) shows the evolution of $B$ with temperature, and the inset zooms in on the crossing region. In this case, we find that the \gls*{CDW} transition occurs at $T_\mathrm{cdw} \approx 1.76|U_\mathrm{eff}|$.

\begin{figure}[t]
\includegraphics[width=\columnwidth]{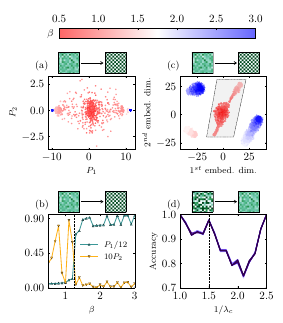}
\caption{Sample results from the three different machine learning methods. \gls*{PCA} (a)-(b) and \gls*{t-SNE} (c) results are shown at $1/\lambda=2$, while \gls*{LBC} (d) results are shown at $\beta=0.30$. (a) The projection of snapshots across all temperatures into the two dimensional space of the first two principal components in the \gls*{PCA}. At extreme temperatures, the density snapshots are similar, and their corresponding data points cluster in the low-dimensional space. Note that at low temperatures, the $\mathbb{Z}_2$ symmetry of the ordered phase is reflected in having two clusters on either side of the high-temperature cluster in the middle. The variances of data points rapidly grow at the critical temperature. This behavior is shown in panel (b), where the average first and second principal components are plotted as a function of $\beta$. (c) Output of the \gls*{t-SNE} applied to the same data shows a similar bunching of points into distinct clusters, which we interpret similarly as belonging to the phases below and above $T_\mathrm{cdw}$. (d) The characteristic $\mathcal{W}$ shape of the accuracy as a function of the guess for the critical inverse \gls*{eph} coupling $1/\lambda_c$ obtained in the \gls*{LBC} method.}
\label{fig:MLMethods}
\end{figure}

\section{Machine Learning Analysis}\label{sec:lbc}

Over the last decade, machine learning and artificial intelligence algorithms have proven to be a powerful complement to more traditional computational methods for the detection of critical phenomena in both classical and quantum many-body systems~\cite{carrasquilla2020machine,johnston2022perspective,dawid2022modern}. In this section, we apply \gls*{PCA}, \gls*{t-SNE}, \gls*{LBC}, and another unsupervised method~\cite{van2017learning, greplova2020unsupervised, gavreev2022learning, richter2023learning, broecker2017quantum} to the bond-Holstein model. In addition to a quantitative confirmation of the \gls*{CDW} phase diagram, the \gls*{LBC} method and the approach of Ref.~\cite{broecker2017quantum} provide us with additional information about the crossover from a metal to a liquid of bipolarons for $T > T_\mathrm{cdw}$ with increasing \gls*{eph} coupling, as was recently observed for the half-filled site-Holstein model~\cite{issa2025learning}.

\subsection{Principal component analysis}
\gls*{PCA} provides a complementary, unsupervised means of identifying structural changes in the configurational data without the need for any physics input. The input to \gls*{PCA} consists of \gls*{DQMC}-generated snapshots of the fermionic densities, collected across a range of parameter values such as the temperature. \gls*{PCA} begins by reshaping each snapshot into a high-dimensional vector of length $N = L\times L$ and constructing the covariance matrix over the full dataset. The eigenvectors of this matrix define orthogonal directions (principal components) that capture the directions of largest variance in the data. The corresponding eigenvalues quantify the contribution of each component to the total variance.

Near a phase transition, the structure of the snapshots changes markedly, and this reorganization manifests as a sharp variation in the leading principal components [e.g., see Fig.~\ref{fig:MLMethods}(a)]. In particular, the first principal component often exhibits an inflection or rapid growth near the transition [exemplified in Fig.~\ref{fig:MLMethods}(b)] reflecting the onset of long-range order. Tracking the magnitude of the first component as a function of the control parameter, therefore, reveals a pronounced feature at the critical point. Consequently, \gls*{PCA} furnishes an unsupervised indicator of $T_\mathrm{cdw}$, with the transition scale identified from the characteristic nonanalytic behavior in the dominant component.

\subsection{t-distributed stochastic neighbor embedding}
\gls*{t-SNE} offers another unsupervised, but (unlike \gls*{PCA}) nonlinear, visualization technique that reveals how the local structure of the configurational data evolves across the phase space. As in the preceding analyses, the input consists of \gls*{DQMC}-generated snapshots of fermionic densities sampled over a range of temperatures. \gls*{t-SNE} operates by constructing a probability distribution over pairwise distances in the high-dimensional snapshot space, such that nearby configurations carry high similarity weight while distant configurations contribute negligibly. It then seeks a \gls*{2D} embedding that preserves these local similarity relationships as faithfully as possible.

The organization of configurations in this reduced space changes qualitatively across a phase transition: snapshots within the same phase cluster tightly, while those near the transition region exhibit increased overlap or form elongated transitional manifolds [see, for example, Fig.~\ref{fig:MLMethods}(c)]. As a result, the arrangement of points in the \gls*{t-SNE} map undergoes a smooth but identifiable reorganization as the control parameter is varied. By tracking the parameter values at which the clustering pattern begins to separate or merge, one obtains a finite interval that reflects the broadened, crossover-like nature of the transition in finite systems. This interval serves as the \gls*{t-SNE}-based estimate of the characteristic transition scale.

\subsection{Learning by confusion}
\gls*{LBC} utilizes configurational snapshots, which in the case of the bond-Holstein model are the fermionic densities, over a particular parameter range. The snapshots are generated with \gls*{DQMC} and acquired across a range of parameter values (e.g., temperature $T$ or \gls*{eph} coupling $\lambda$). A critical point is then chosen, dividing the samples into two groups based on whether their parameter value is above or below the critical value. A \gls*{CNN} is then trained to learn the classification. If the critical point is chosen to be at either of the extremes of the range, all snapshots share the same label, and the \gls*{CNN} can easily learn to return that value with nearly perfect accuracy. On the other hand, if the critical point is chosen in the middle of the range but at an incorrect location, the \gls*{CNN} is `confused' since mislabeled snapshots do not conform to their designated phase. Its accuracy is low in this case. However, if the critical point is correct, the \gls*{CNN}'s accuracy improves again. As a consequence, a plot of the accuracy as a function of the parameter has a characteristic ${\cal W}$ shape [e.g., Fig.~\ref{fig:MLMethods}(d)] - high at the `trivial' extremes of range, with an interior maximum that identifies the actual critical point.

\subsection{Another unsupervised neural network method}
Similarly to \gls*{LBC}, the method by Broecker, Assaad, and Trebst in Ref.~\cite{broecker2017quantum} utilizes a series of supervised trainings using a neural network with a critical point chosen at different points in a range of interest. But unlike in \gls*{LBC}, the data used for each training is always a balanced mix of \gls*{DQMC} fermion density snapshots at the two sides of the chosen critical point. A plot of the training accuracy as a function of the tuning parameter will therefore have a dominant peak at the true critical point.

\begin{figure}[t]
\includegraphics[width=\columnwidth]{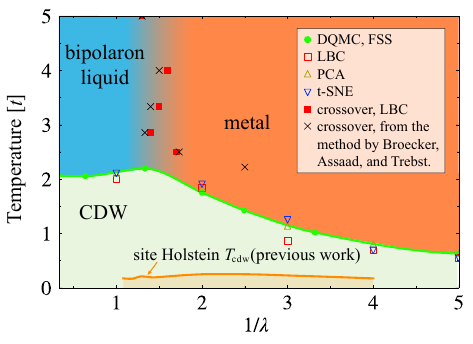}
\caption{The phase diagram for the half-filled bond-Holstein model on a square lattice as a function of the inverse dimensionless coupling constant $\lambda^{-1}$.
For comparison, the green (orange) lines show the value of $T_\mathrm{cdw}$ for the bond (site) model obtained
from a finite-size scaling analysis of the charge structure factor (e.g., see Fig.~\ref{fig:ScaledScdw}). The remaining data points indicate estimates obtained from different machine Learning methods, including learning by confusion (LBC, red $\square$), principal component analysis (PCA, yellow $\triangle$), \gls*{t-SNE} (blue $\triangledown$), and the method introduced in Ref.~\cite{broecker2017quantum} (black $\times$). LBC, PCA, and t-SNE accurately identify $T_\mathrm{cdw}$, the temperature below which a CDW regime is observed. \gls*{LBC} and the method of Ref.~\cite{broecker2017quantum} additionally provide an estimate for the crossover line separating the metal and bipolaron liquid regimes. 
The transition temperatures for the site-Holstein model were taken from Ref.~\cite{issa2025learning}. Bond-Holstein simulations for $1/\lambda_\mathrm{bond}<0.5$ are difficult to equilibrate. Nevertheless, the decrease in $T_\mathrm{cdw}$ in the strong coupling limit (small $1/\lambda$) observed for the site-Holstein model is absent for the bond model.
}
\label{fig:phasediagram}
\end{figure}

\section{Phase Diagram}\label{sec:phase_diagram}

Figure \ref{fig:phasediagram} shows the $(1/\lambda,T)$ phase diagram of the half-filled bond-Holstein model, which is the main result of this work. For reference, the relative \gls*{CDW} phase boundary for the equivalent site-Holstein model, obtained from Ref.~\cite{issa2025learning}, is also shown. In this case, the solid lines indicate the \gls*{CDW} transition temperatures obtained from a finite size scaling analysis of the \gls*{DQMC} data similar to that presented in Fig.~\ref{fig:ScaledScdw}. The data presented in that figure was for $\lambda_\mathrm{bond} = 0.4$ ($1/\lambda_\mathrm{bond}=2.5$), where $T_\mathrm{cdw}^\mathrm{bond} \approx 3T_\mathrm{cdw}^\mathrm{site}$. This enhancement is even larger for increased coupling strength. For example, $T_\mathrm{cdw}^\mathrm{bond}$ is an order of magnitude larger than its site counterpart at the smallest $1/\lambda_\mathrm{bond}$ value for which we have data. Strong coupling brings out this effect most dramatically where, as we argued earlier, $T_{\rm cdw} \sim \lambda$ for bond model while $T_{\rm cdw} \sim 1/\lambda$ for the site model.

Figure~\ref{fig:phasediagram} also shows our $T_\mathrm{cdw}$ estimates obtained from the \gls*{LBC}, \gls*{PCA}, \gls*{t-SNE}, and the method by Ref. \cite{broecker2017quantum}. For each fixed $\lambda$, 100 fermion density snapshots per temperature were used by \gls*{LBC}, \gls*{PCA}, and \gls*{t-SNE} to detect the \gls*{CDW} phase transition. The temperature grid was chosen independently for each lambda, with temperatures spanning the range $0.50 \le T \le 3.0$. All three machine learning approaches accurately capture the \gls*{CDW} transition and provide complementary, model-agnostic indicators of the phase boundary.

In addition to successfully locating the \gls*{CDW} transition, \gls*{LBC} also reveals a high-temperature crossover between a metallic regime at weak coupling and a liquid of localized bipolarons at strong coupling~\cite{issa2025learning}. Importantly, the method by Ref. \cite{broecker2017quantum} also detects the same crossover region with results that are consistent with \gls*{LBC}. Note, however, that there exists one data point at $\beta=0.45$ that was not resolved by \gls*{LBC}, where the method of Ref.~\cite{broecker2017quantum} predicts $1/\lambda_c=2.50$. If this point is included, it suggests there is a sharp turn in the crossover line towards the right of the phase diagram. We speculate that \gls*{LBC} failed at this $\beta$ because the training is performed along a line in parameter space that is fairly too close to the \gls*{CDW} phase boundary. 

When determining the crossover points with \gls*{LBC} and the method of Ref.~\cite{issa2025learning}, we used 1000 snapshots per $\lambda$ value on a grid of sixteen values spanning $1/\lambda \in [1, 2.5]$ at every fixed temperature. The crossover, reflected in the position and curvature of the \gls*{LBC} decision boundary, marks the temperature scale at which charge carriers transition from being delocalized to tightly bound electron and phonon composites~\cite{issa2025learning}. Together, these machine-learning approaches yield a consistent and physically transparent picture of both the \gls*{CDW} transition and the underlying evolution of charge dynamics across the phase diagram.

\section{Summary \& Discussion}\label{sec:conclusions}

We have studied the \gls*{CDW} transition of the half-filled bond-Holstein model using a combination of \gls*{DQMC} and machine learning methods, and compared the model's ordering tendencies to those previously observed for the site-Holstein model. 
We found that the bond-Holstein model exhibits stronger tendencies toward \gls*{CDW} order and bipolaron formation. In particular, the critical temperatures for the bond model are significantly higher than those in the site model for a given value of the dimensionless coupling $\lambda$. We also found that the two models behave differently in the strong coupling (large $\lambda$) limit. Unlike the behavior of its more commonly studied site counterpart, where $T_\mathrm{cdw}^\mathrm{site}$ monotonically decreases with $\lambda$ in the strong coupling limit, the transition temperature for the bond-Holstein model increases with $\lambda$ and has a finite value in the $\lambda\rightarrow \infty$ (or equivalently $t = 0$) limit. 
While it is unclear whether the bond-Holstein model continues to increase or plateaus at large $\lambda$ due to computational difficulties (increasing autocorrelation times) at large coupling, the stark difference between the two models is nonetheless noteworthy.

As mentioned earlier, the bond-Holstein model can be regarded as a low-energy single-band description of oxygen-bond-stretching modes in oxides like the cuprates and bismuthates Ba$_{1-x}$K$_x$BiO$_3$. The latter has a very robust \gls*{CDW} phase that has been understood previously using a multi-orbital model with Bi-O bond-stretching phonons~\cite{CohenStead2023hybrid}. Our results suggest that an effective single-band description using the bond-Holstein model may also be possible.

An obvious and very interesting extension of our work would be to study the doped bond-Holstein model, and especially the prospects for superconductivity. This would be especially timely given the known difficulty (very low transition temperatures) in accessing the \gls*{SC} phase in the site-Holstein model~\cite{bradley21}. Given that \gls*{CDW} and bipolaron formation, enemies of \gls*{SC}, are more robust in the bond-Hamiltonian, higher \gls*{SC} response seems at first glance unlikely. In this light, it would also be interesting to carry out a detailed comparison of \gls*{DQMC} simulations for the bond-Holstein model with predictions obtained from Eliashberg theory similar to those previously carried out for the site-Holstein case~\cite{esterlis18}. There it was noted that the Eliashberg prediction that the superconducting $T_\mathrm{c}$ increases monotonically with the \gls*{eph} coupling is fatally flawed at large $\lambda$ due to polaron formation that is properly captured by \gls*{DQMC}. It would be interesting to explore when and where similar breakdowns of Eliashberg theory occur for the bond-Holstein model, with implications for understanding pairing in different oxide superconductors.

\begin{acknowledgments}
This work was supported by the grant DE-SC0022311 funded by the U.S. Department of Energy, Office of Science.
\end{acknowledgments}

\bibliography{bondholstein}

\end{document}